\documentstyle[aps]{revtex}

\begin{document}
\draft
\title{Trapped atomic condensates with anisotropic interactions}
\author{S. Yi and L. You}
\address{School of Physics, Georgia Institute of Technology,
Atlanta, GA 30332-0430}
\date{\today}
\maketitle

\begin{abstract}
We study the ground state properties of trapped atomic condensates
with electric field induced dipole-dipole interactions.
A rigorous method for constructing the pseudo potential in the spirit
of ladder approximation is developed for general non-spherical (polarized)
particles interacting anisotropically.
We discuss interesting features not previously considered for
currently available alkali condensates. In addition
to provide a quantitative assessment for controlling atomic
interactions with electric fields, our investigation may also
shed new light into the macroscopic coherence properties of the
Bose-Einstein condensation (BEC) of dilute interacting atoms.
\end{abstract}


\narrowtext

The success of atomic Bose-Einstein condensation (BEC)
\cite{bec,mit,rice,Edwards} has stimulated great interest
in the properties of trapped quantum gases.
In standard treatments of interacting quantum gases,
realistic inter-atomic potentials $V(\vec R)$ are replaced by
contact forms $u_0\delta(\vec R)$ in the so-called
shape independent approximation (SIA) \cite{yang}.
Such an approximation results in tremendous simplification.
To date, the SIA has worked remarkably well as recent
theoretical investigations \cite{Edwards} have successfully
accounted for almost all experimental observations \cite{glauber,esry}.

Currently available degenerate quantum gases are cold and dilute,
with interactions dominated by low energy binary collisions.
When realistic interatomic potentials are
assumed to be isotropic and short ranged, i.e. decreasing faster
than $-1/R^3$ asymptotically for large interatomic separations $R$,
the properties of
a complete two body collision is described by just one
atomic parameter: $a_{\rm sc}$,
the s-wave {\it scattering length}.
The scattering amplitude is isotropic
and energy-independent:
$f(\vec k,\vec k')=-4\pi a_{\rm sc}$ for collisions involving
incident momentum $\vec k$ scattering into $\vec k'$.
Effective physical mechanisms exist for control of
the atom scattering lengths \cite{gora,verhaar,eite}.
If implemented, these control
`knobs' allow for unprecedented comparison between
theory and experiment over a wide range of interaction strength.
Indeed, very recently several groups have successfully
implemented {\it Feshbach resonance} \cite{fesh},
thus enabling a control knob on $a_{\rm sc}$ through the
changing of an external magnetic field.
Other physical mechanisms also exist for modifying
atom-atom interactions, e.g. the {\it shape resonance}
due to anisotropic dipole interactions
inside an external electric field \cite{mm}.

Although fermions with anisotropic interactions are
well studied within the context of $^3$He fluid \cite{leggett}
and in d-wave high $T_c$ superconductors, anisotropically
interacting bosons have not been studied in great detail.
In particular, we are not aware of any systematic approach
for constructing an anistropic pseudo potential \cite{yang}.

In this paper, we study the ground state properties of
trapped condensates with dipole interactions.
A rigorous method is developed for constructing the
anisotropic pseudo potential that can also be applied
to future polar molecular BEC \cite{doyle,stwalley}.
This Letter is
organized as follows. First we briefly review the
SIA pseudo-potential approximation.
We then construct an analogous effective low energy
anisotropic pseudo-potential.
Numerical results are then discussed for
$^{87}$Rb \cite{bec} inside the external E-field in
the JILA TOP trap. We conclude with a brief discussion
of prospects for realistic experiments.

For $N$ trapped spinless bosonic atoms in a potential $V_t(\vec r)$,
the second quantized Hamiltonian is given by
\begin{eqnarray}
{\cal H}&=& \int\! d\vec r\,\hat\Psi^{\dag}(\vec r)
\left[-\frac{\hbar^2}{2M}\nabla^2+V_{t}(\vec r)
-\mu\right]\hat\Psi(\vec r) \nonumber\\
&\ &+\frac{1}{2}\int\! d\vec r\int\! d\vec r'
\hat\Psi^{\dag}(\vec r)\hat\Psi^{\dag}(\vec r')
V(\vec r-\vec r')\hat\Psi(\vec r')\hat\Psi(\vec r),
\label{h}
\end {eqnarray}
where $\hat\Psi(\vec r)$ and $\hat\Psi^{\dag}(\vec r)$ are atomic
(bosonic) annihilation and creation fields.
The chemical potential $\mu$ guarantees the
atomic number $\hat N=
\int d\vec r\,\hat\Psi^{\dag}(\vec r)\hat\Psi(\vec r)$
conservation.

The bare potential $V(\vec R)$ in (\ref{h})
needs to be renormalized for a meaningful
perturbation calculation. For bosons, the
usual treatment is based on field theory and
is rather involved \cite{yang,Galits,Beliaev,Brueckner}.
Physically the SIA can be viewed as a valid low energy and low density
renormalization scheme. The physics involved is rather simple:
one simply replaces the bare potential
$V(\vec R)$  by the pseudo potential
$u_0\delta(\vec R)$ such that whose
first order Born scattering amplitude
reproduces the complete scattering amplitude
($-a_{\rm sc}$). This requires
$u_0=4\pi\hbar^2 a_{\rm sc}/M$.

When an electric field is introduced along the positive z
axis, an additional dipole interaction
\begin{eqnarray}
V_{E}(\vec R)&&=-u_2{Y_{20}(\hat R)\over R^3},
\label{ve}
\end{eqnarray}
appears, where $u_2=4\sqrt{(\pi/5)}\,\alpha(0)\alpha^*(0){\cal E}^2$,
with $\alpha(0)$ being the polarizability,
and ${\cal E}$ the electric field strength.
As was shown in Ref. \cite{mm},
this modification
results in a completely new low-energy scattering
amplitude
\begin{eqnarray}
f(\vec k,\vec k')\Big|_{k=k'\to 0}=4\pi\sum_{lm,l'm'}
t_{lm}^{l'm'}({\cal E})Y_{lm}^*({\hat k})Y_{l'm'}({\hat k'}),
\label{cf}
\end{eqnarray}
with $t_{lm}^{l'm'}({\cal E})$ the reduced T-matrix
elements. They are all energy independent
and act as generalized scattering lengths.
The anisotropic $V_{E}$ causes the dependence on both
incident and scattered directions: $\hat k$ and $\hat k'=\hat R$.

A general anisotropic pseudo potential can be
constructed according to
\begin{eqnarray}
V_{\rm eff}(\vec R)=u_0\delta(\vec R)
+\sum_{l_1>0, m_1}  \gamma_{l_1m_1}
{Y_{l_1m_1}(\hat R)\over R^3},
\label{veff}
\end{eqnarray}
whose first Born amplitude is then given by
\begin{eqnarray}
f_{\rm Born}(\vec k,\vec k')
=-(4\pi)^2 a_{\rm sc}Y_{00}^*(\hat k)Y_{00}(\hat k')
-{M\over 4\pi\hbar^2}\sum_{l_1 m_1} \gamma_{l_1m_1} (4\pi)^2 \sum_{lm}
\sum_{l'm'}{\cal T}_{lm}^{l'm'}(l_1,m_1)
Y_{lm}^*(\hat k)Y_{l'm'}(\hat k'),
\label{bf}
\end{eqnarray}
with $
{\cal T}_{lm}^{l'm'}(l_1,m_1)=(i)^{l+l'}{\cal R}_{l}^{l'}
I_{lm}^{l'm'}(l_1,m_1).
$
Both
\begin{eqnarray}
I_{lm}^{l'm'}(l_1m_1)&&=\langle
Y_{l'm'}|Y_{l_1m_1}|Y_{lm}\rangle, {\hskip 24pt \rm and}\nonumber\\
{\cal R}_{l}^{l'} &&=\int_0^{\infty} d R\,{1\over
R}j_l(kR)j_{l'}(k'R),\nonumber
\end{eqnarray}
can be computed analytically \cite{su}. The $1/R^3$
form in Eq. (\ref{veff}) assures all ${\cal R}_{l}^{l'}$ to be
$k=k'$ independent [by a change of variable to
$x=kR$ in the integral]. Putting
\begin{eqnarray}
f_{\rm Born}(\vec k,\vec k')=f(\vec k,\vec k'),
\end{eqnarray}
one can solve for the $\gamma_{l_1m_1}({\cal E})$
as $t_{lm}^{l'm'}({\cal E})$ are known numerically \cite{mm,mm2}.
This reduces to the linear equations
\begin{eqnarray}
-{M\over 4\pi \hbar^2}\sum_{l_1 m_1} \gamma_{l_1m_1} (4\pi)
{\cal T}_{lm}^{l'm'}(l_1,m_1)\equiv
t_{lm}^{l'm'},
\end{eqnarray}
for all ($lm$) and ($l'm'$) with $l,l'\ne 0$,
and separately $a_{\rm sc}({\cal E})=-t_{00}^{00}({\cal E})$.
The problem
simplifies further for Bosons (fermions) as only
even (odd) $(l,l')$ terms are needed to match.
Figure \ref{fig1} displays result of
$a_{\rm sc}({\cal E})$ for the triplet state of $^{87}$Rb.
The Born amplitude for the dipole term $V_E$ is
\begin{eqnarray}
f_{\rm Born}(\vec k,\vec k') &&=u_2{M\over 4\pi \hbar^2}
(4\pi)^2{\cal T}_{00}^{20}
\sum_{lm,l'm'}\overline{\cal T}_{lm}^{l'm'}
Y_{lm}^*(\hat k)Y_{l'm'}(\hat k'),
\end{eqnarray}
with
${\cal T}_{00}^{20}=-0.023508$.
$\overline {\cal T}_{lm}^{l'm'}={\cal T}_{lm}^{l'm'}(2,0)/{\cal T}_{00}^{20}$
are tabulated below for small $(l,l')$.
\begin{table}
\caption{$\bar{\cal T}_{lm}^{l'm'}$}
\begin{tabular}{c|ccccc}
$(lm),(l'm')$&(00)&(20)&(40)&(60)&(80)\\ \hline (00)&0&1&0&0&0\\
(20)&1&-0.63889&0.14287&0&0\\(40)&0&0.14287&-0.17420&0.05637&0\\
(60)&0&0&0.05637&-0.08131&0.03008\\(80)&0&0&0&0.03008&-0.04707
\end{tabular}
\label{table1}
\end{table}
We found that away from {\it shape resonances},
Table (\ref{table1}) agrees ($\sim$
a few per cent) with the same ratios
$t_{lm}^{l'm'}({\cal E})/t_{00}^{20}({\cal E})$
from the numerical multi-channel calculations \cite{mm}.
This interesting observation applies for all
bosonic alkali triplet states we computed:
$^7$Li, $^{39,41}$K, and $^{85,87}$Rb,
for up to a field strength of $3\times 10^6$ (V/cm) \cite{mm,mm2}.
Physically, this implies that effect of $V_E$
is perturbative as ${\cal E}$
remains small in atomic units.
What is remarkable is that
${\cal T}_{00}^{20}({\cal E})$ and $t_{00}^{20}({\cal E})$
also agree in absolute values \cite{mm}.
For $^{87}$Rb, we found
\begin{eqnarray}
u_2{M\over 4\pi\hbar^2}
(4\pi)^2{\cal T}_{00}^{20}
=-1.495\times 10^{10} {\overline {\cal E}}^2 (a_0),
\end{eqnarray}
with ${\overline {\cal E}}$ in
atomic units ($5.142\times 10^9$ V/cm).
$a_0$ is the Bohr radius. While multi-channel scattering gives \cite{mm2}
\begin{eqnarray}
(4\pi)t_{00}^{20} =-1.512\times 10^{10} {\overline {\cal E}}^2 (a_0).
\end{eqnarray}
The cause of this slight difference (1\%)
is not entirely clear and but is within numerical error.

We can thus approximate Eq. (\ref{veff}) by keeping only
the $l_1=2,m_1=0$ term in the sum
\begin{eqnarray}
V_{\rm eff}(\vec R)=u_0\delta(\vec R)-u_2 Y_{20}(\hat R)/R^3,
\end{eqnarray}
away from the {\it shape resonance}.
At zero temperature the condensate wave function
$\psi(\vec r,t)=\langle\hat\Psi(\vec r,t)\rangle$ then obeys the
following nonlinear Schrodinger equation
\begin{eqnarray}
i\hbar {d\over dt}\psi(\vec r,t)
&&=\left[-\frac{\hbar^2}{2M}\nabla^2+V_{t}(\vec r)
-\mu+u_0|\psi(\vec r,t)|^2\right.\nonumber\\
&&\left.-u_2\int d\vec r' {Y_{20}(\hat R)\over R^3}
|\psi(\vec r',t)|^2\right]\psi(\vec r,t),
\label{nlse}
\end{eqnarray}
with $\psi(\vec r,t)$ normalized to $N$. The ground state
is found by steepest descent through propagation
of Eq. (\ref{nlse}) in imaginary time $(it)$.
For a cylindrical symmetric trap
$V_{\rm t}(\vec r)=M(\omega _\perp^2x^2+\omega_\perp^2y^2+\omega _z^2z^2)/2$,
the ground state also possesses azimuthal symmetry. Therefore
the non-local term simplifies to
\begin{eqnarray}
\int d\vec r'|\psi(\rho',z')|^2 {Y_{20}(\hat R)\over
R^3}=\int dz' d\rho' {\cal K}(.,.;.)
|\psi(\rho',z')|^2,\nonumber
\end{eqnarray}
with the kernel ${\cal K}(\rho,\rho';z-z')$
expressed in terms of the standard Elliptical
integrals ${\rm E}[.]$ and ${\rm K}[.]$. The kernel is
divergent at $\vec r=\vec r'$, so a cut-off radius $R_c$
is chosen such that ${\cal K}(\rho',\rho,z'-z)=0$
whenever $|\vec r-\vec r'|<R_c$. We typically
$R_c\sim 50 (a_0)$, much smaller than the grid size,
to minimize numerical errors.
Technical details for numerical computations and
for handling the singular
rapid variation of the kernel over
small length scale will be discussed elsewhere \cite{su}.

Figure \ref{fig2} presents $\psi(\rho,z)$ along
$\rho=0$ (a) and $z=0$ (b) cuts respectively for $^{87}$Rb
($a_{\rm sc}=5.4$ nm) at several different ${\cal E}$. We note
the condensate shrinks radially while stretches along z-axis
 to minimize the
dipole interaction $V_E$. The top-right corner inserts shows
electric field polarized atoms in (radially) repulsive (a) and
(longitudinally) attractive (b) configurations.
An elongated condensate along the z-axis
reduces the total energy. The same mechanism could cause
spontaneous alignment of polar molecular condensates inside
isotropic traps \cite{doyle}.
For better insights we try a variation ansatz
\begin{eqnarray}
\psi_T(\rho,z)={\kappa^{1/2}\over \pi^{3/4}d^{3/2}} \exp\left[
-{1\over 2d^2}(\rho^2+\kappa^2 z^2)\right],
\end{eqnarray}
with parameters $d$ and $\kappa$. In
dimensionless units for length
($a_{\perp}=\sqrt{\hbar/M\omega_\perp}$), energy
($\hbar\omega_\perp$), and $\lambda=\omega_z/\omega_\perp$, we
obtain
\begin{eqnarray}
E[\psi_T]&&=(1+{\lambda^2\over 2\kappa^2})d^2+(1+{\kappa^2\over
2}){1\over d^2} +{4N\kappa\over\sqrt{2\pi}} {a_{\rm sc}^{\rm
eff}\over a_{\perp}}{1\over d^3}, \label{energy}
\end{eqnarray}
with the effective scattering length
$a_{\rm sc}^{\rm eff}=a_{\rm sc}[1-b(\kappa){u_2/
u_0}]$, and
\begin{eqnarray}
b(\kappa)={\sqrt{5\pi}\over
3(\kappa^2-1)}\left(-2\kappa^2-1+{3\kappa^2\tanh^{-1}\sqrt
{1-\kappa^2}\over \sqrt {1-\kappa^2} }\right).\nonumber
\end{eqnarray}
The $b(\kappa)$ is monotonically decreasing, and bounded between
$b(0)=\sqrt{5\pi}/3$ and $b(\infty)=-2\sqrt{5\pi}/3$.
$a_{\rm sc}^{\rm eff}$ is shown in
Fig. \ref{fig3} as a function of ${\cal E}$
for several different values of $\kappa$.
For increasing electric field ${\cal E}$, variational calculation
results in decreasing $\kappa$, eventually $\kappa$ becomes less
than one, i.e. the condensate changes from oblate (pancake) shaped
at zero field (for the TOP trap) to prolate (cigar) shaped. We
also note that $b(\kappa)\ge 0$ for $\kappa\le 1$, therefore
$a_{\rm sc}^{\rm eff}$ becomes negative at certain field value
${\cal E}_c$ in the case of a positive $a_{\rm sc}({\cal E}=0)$,
causing the collapse of the condensate. This is indeed what we
found as illustrated in Fig. \ref{fig4}. A detailed discussion of
the collapse and other interesting
features will be given elsewhere \cite{su}.

We note the energy of dipole alignment
\begin{eqnarray}
E_P &&\sim -(2\pi) 1\times10^{18}
\times {\overline{\cal E}}^2 {\rm (Hz)},
\end{eqnarray}
becomes much larger than the trap depth
at the proposed ${\cal E}$ values for $^{87}$Rb.
Therefore spatial homogeneity for ${\cal E}(\vec r)$ is required.
At ${\cal E}\sim 5\times 10^5$ (V/cm) [${\overline{\cal E}}\sim 10^{-4}$]
with a spatial gradient
$<10^{-4}$/cm$^3$, the corresponding force
is smaller than the magnetic trapping force
for typical traps at $\sim 100$ (Hz).
For comparison, the magnetic field gradient is
$\sim 10^{-6}$/cm$^3$ inside the Penning trap magnets.
Although the proposed electric field [$10^5$ (V/cm) $< {\cal E}<10^6$ (V/cm)]
is large, it can be created through careful laboratory techniques as breakup is
fundamentally limited by field ionization, which
typically occurs at ${\cal E}>10^7$ (V/cm) \cite{latham}.
Recently a ${\cal E}$ field of upto $1.25\times 10^5$ (V/cm)
was used successfully to decelerate a molecular beam \cite{meijer}.

In conclusion, we have developed a general scheme for
constructing effective pseudo-potentials for anisotropic
interactions. Our scheme guarantees that the first order
Born scattering amplitude from the pseudo-potential
reproduces the complete
scattering amplitude obtained from a multi-channel
computation including the anisotropic dipole-interaction,
thus contains no energy dependence at low temperatures
of the trapped atomic gases \cite{mm}. Our scheme is
thus more pleasing than the standard
Skyrme type velocity dependent effective potentials
commonly adopted in nuclear physics \cite{ring}.
We also presented results for both the electric field
modified atomic scattering parameters and the
induced changes to the condensate
for $^{87}$Rb in the JILA TOP trap. Our theory can be
directly extended to systems involving magnetic dipole interaction
of atoms/molecules in a static magnetic trap and systems
of trapped molecules with permanent electric dipoles \cite{doyle}.
For alkali atoms, typical magnetic dipole interaction is weak
since a Bohr magneton ($\mu_B=e\hbar/2mc$) only corresponds to
an electric dipole of $\sim (1/2\alpha_f)(e a_0)$ (fine structure constant
$\alpha_f\approx 1/137$), which is equivalent to the
induced electric dipole at ${\cal E}=6\times 10^4$ (V/cm)
for $^{87}$Rb. Other atoms with
larger magnetic dipole moments \cite{doyle2} will
display clearer anisotropic effects. Typical
hetero-nuclear diatomic molecules have a permanent
electric dipole moment of $\sim (e a_0)$, corresponding to
an induced moment in $^{87}$Rb at ${\cal E}=1.6\times 10^7$ (V/cm)
\cite{doyle}. Trapped molecules with aligned permanent
electric dipoles (by an external E-field) would
give similar results. However, magnetic trapped molecules \cite{doyle}
with unaligned electric dipoles interacting with the spin axis
represents an interesting
extension that requires further investigation.

We thank Dr. M. Marinescu for helpful discussions during the
early stages of this work.
This work is supported by the U.S. Office of Naval Research
grant No. 14-97-1-0633 and by the NSF grant No. PHY-9722410.

\begin{figure}
\caption{The field dependent value for $a_{\rm sc}$. Note the
shape resonance for ${\cal E}$ around $8.3\times 10^5$ (V/cm).}
\label{fig1}
\end{figure}

\begin{figure}
\caption{(a) $\psi(\rho,0)$ for $^{87}$Rb with
$\omega_\perp=(2\pi) 70$ (Hz), $\omega_z=\sqrt{8}\,\omega_\perp$,
and $N=5000$ atoms. Solid, dashed-dot, dashed, and dotted
lines are for
${\cal E}=0, 4.0\times 10^5, 5.7\times 10^5$, and
$5.88\times 10^5$ (V/cm) respectively.
$a_{\perp}$ is the radial trap width.
  \\ (b) same as in
Fig. 2 (a), but for $\psi(0,z)$.} \label{fig2}
\end{figure}

\begin{figure}
\caption{Typical behavior of the effective scattering length
$a_{\rm sc}^{\rm eff}.$ Lines corresponds to
$\kappa=5.1,1.7,1.02,0.34,0.017,$ in descending order of
$a_{\rm sc}^{\rm eff}$.}\label{fig3}
\end{figure}

\begin{figure}
\caption{Electric field dependence of the
width aspect ratio for parameters of Fig. 2. The
solid line is the result of our variational calculation
while circles denote exact numerical results.
The dashed line corresponds to
$\sqrt{\sqrt{8}}$, the result for a non-interacting
gas in the TOP trap.}\label{fig4}
\end{figure}

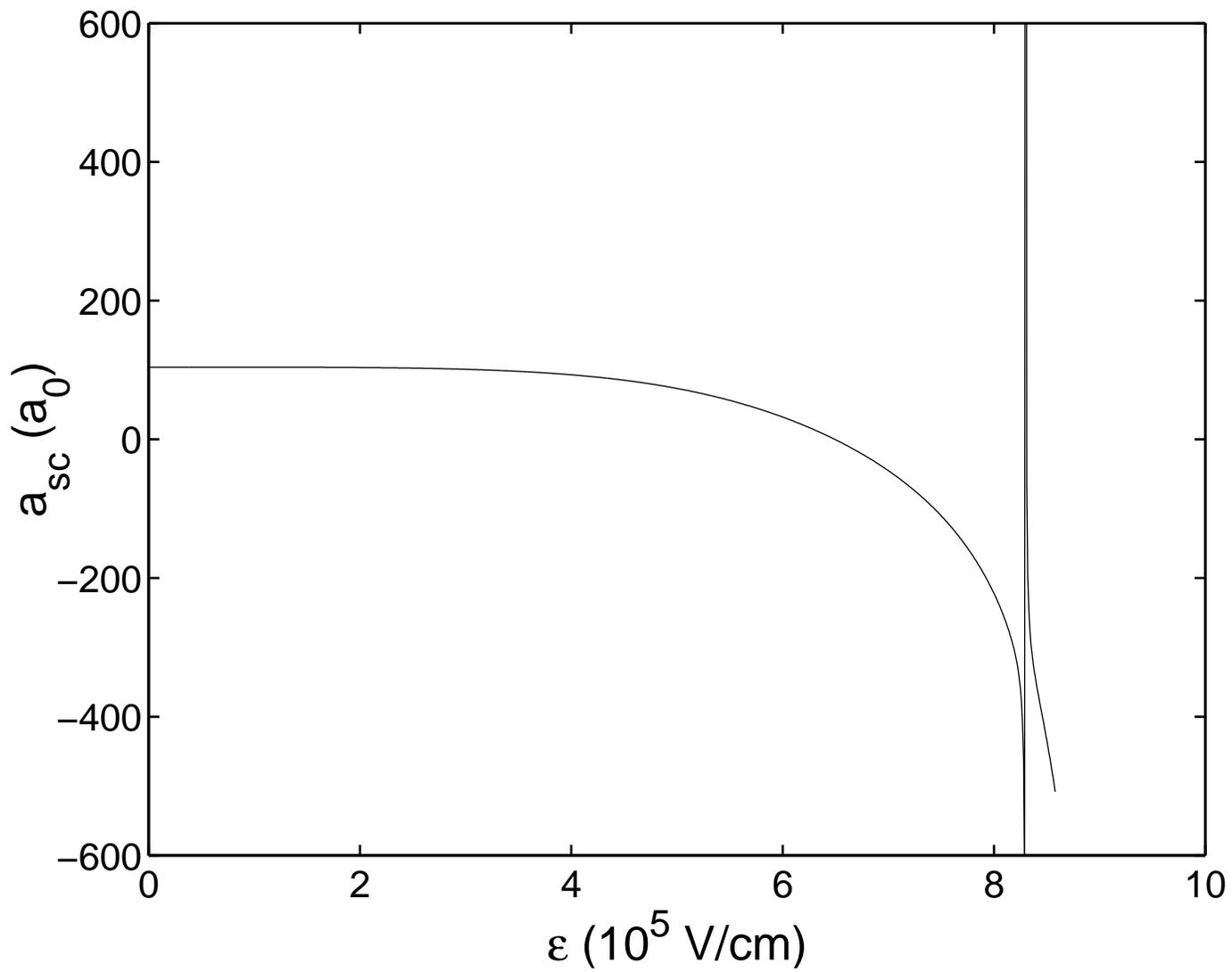
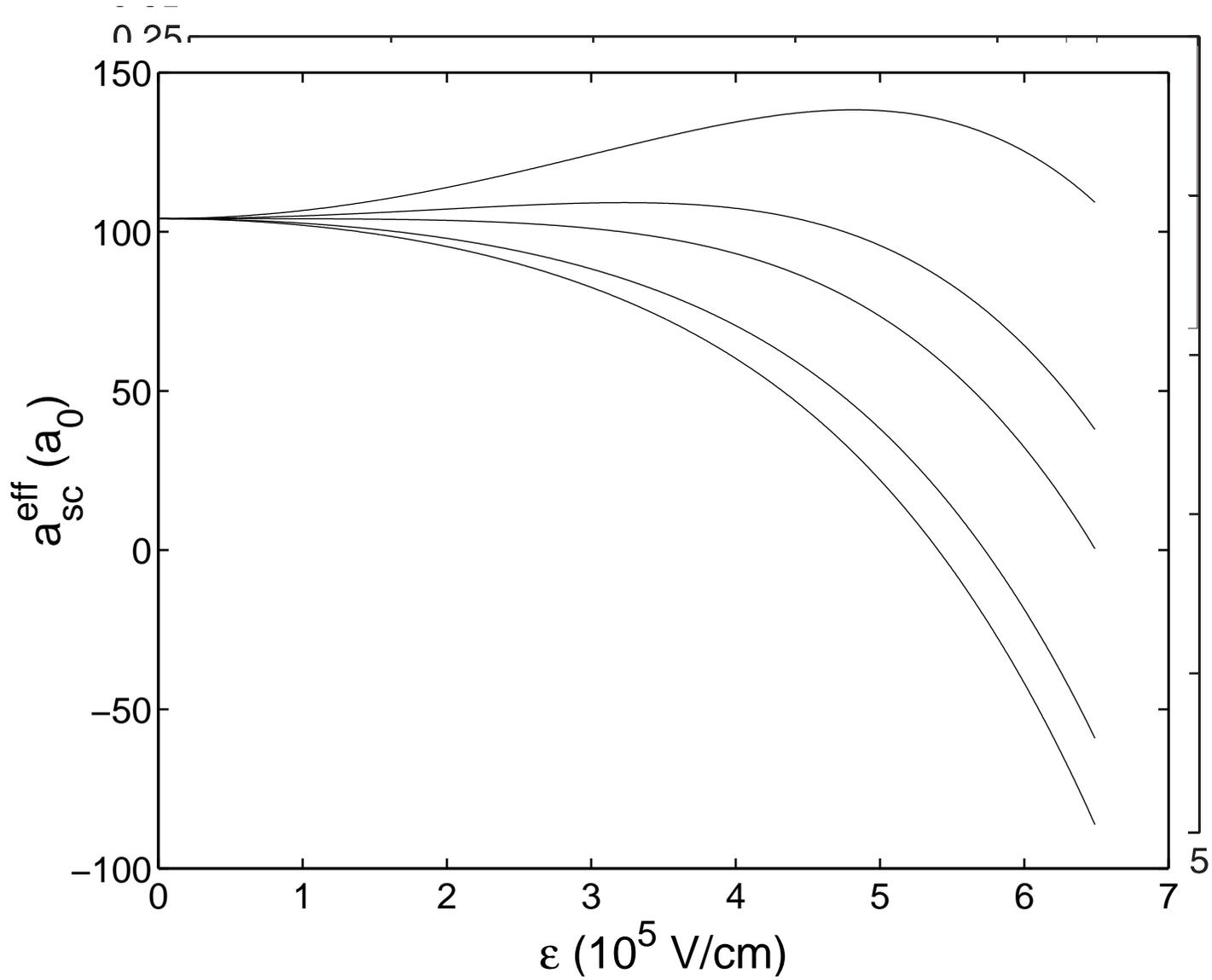
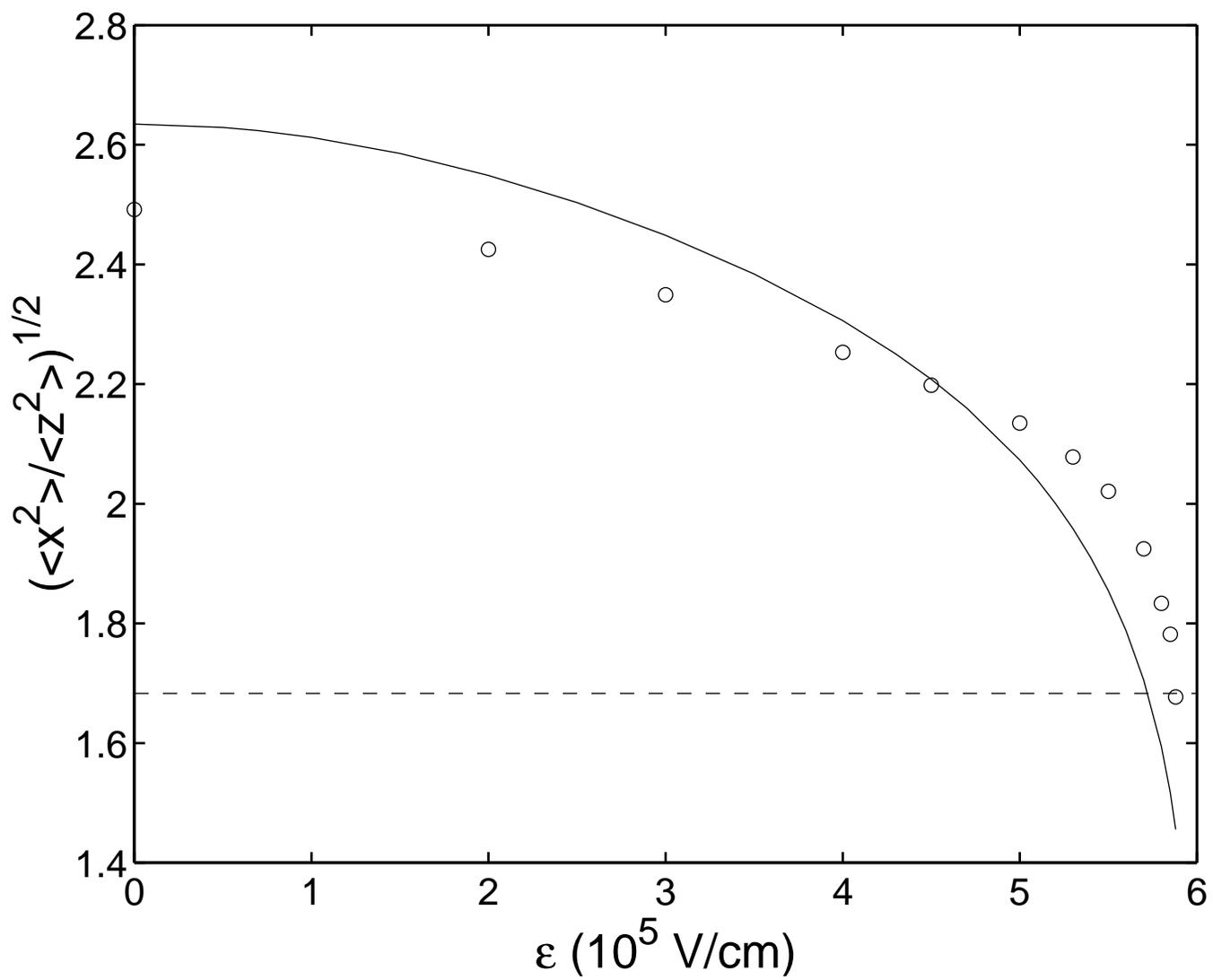
\end{document}